# Nanoelectromechanics of Polarization Switching in Piezoresponse Force Microscopy

Sergei V. Kalinin,[a] A. Gruverman,[b] J. Shin,[a] A.P. Baddorf,[a]

E. Karapetian,[c] and M. Kachanov[c]

[a] Condensed Matter Sciences Division, Oak Ridge National Laboratory, Oak Ridge, TN 37831

[b] Department of Materials Science and Engineering, North Carolina State University, Raleigh, NC 27695

[c] Department of Mechanical Engineering, Tufts University, Medford, MA 02155

## Abstract

Nanoscale polarization switching in ferroelectric materials by Piezoresponse Force Microscopy (PFM) in weak and strong indentation limits is analyzed using exact solutions for electrostatic and coupled electroelastic fields below the tip. It is proposed that the tip-induced domain switching can be mapped on the Landau theory of phase transitions with the domain size as an order parameter. For a point charge interacting with a ferroelectric surface, switching of both first and second order is possible depending on the charge-surface separation. For a realistic tip shape, the domain nucleation process is first order in charge magnitude and polarization switching occurs only above a critical tip bias. In pure ferroelectric or ferroelastic switching, the late stages of the switching process can be described using point charge/force model and arbitrarily large domains can be created; however, the description of the early stages of nucleation process when domain size is comparable with the tip radius of curvature requires exact field structure to be taken into account.

The drive towards nanotechnology necessitates new ways to manipulate the properties of matter at the nanoscale. In the last several years, significant attention has been attracted to application of Piezoresponse Force Microscopy (PFM) for characterization of ferroelectric materials that are used in high-density non-volatile memories and other electronic devices.[1,2,3] PFM provides a novel approach to nanoscale engineering via local modification and control of ferroelectric domain structures with ~10-30 nm resolution.[4,5,6,7] The practical viability of these PFM applications is critically dependent on the minimal stable domain size that can be formed during local polarization switching induced by a tip-generated field. Analysis of domain switching processes using a point charge approximation using the Landauer model for domain geometry has been given by Molotskii *et al.*[8,9] and independently by Abplanalp.[10] However, the point charge model is clearly inapplicable for the description of a realistic tip shape when the tip size is comparable with the domain size. More importantly, the point charge model completely ignores strain effects. It has been shown previously that mechanical strain produced by the tip can suppress local polarization[11] or induce local ferroelectroelastic polarization switching.[12,13] Here, we quantitatively analyze the local polarization switching using exact expressions for the electroelastic fields below the PFM tip in the weak and strong indentation regimes[14] derived elsewhere.[15] It is suggested that the tip-induced polarization switching can be described in the framework of the Landau theory of phase transitions with a domain size as an order parameter.

The driving force for the 180° polarization switching process in ferroelectrics is the change in the bulk free energy density:[10,12]

$$\Delta g_{bulk} = -\Delta P_i E_i - \Delta d_{i\mu} E_i X_\mu, \qquad (1)$$

where $P_i$, $E_i$, $X_\mu$, and $d_{i\mu}$, are the components of the polarization, electric field, stress and piezoelectric constant tensor correspondingly, $i = 1,2,3$, and $\mu = 1,..,6$. The first and the second terms in Eq.(1) describe ferroelectric and ferroelectroelastic switching respectively. The free energy of the nucleating domain is

$$\Delta G = \Delta G_{bulk} + \Delta G_{wall} + \Delta G_{dep},  \qquad (2)$$

where the first term is the volume change in free energy, $\Delta G_{bulk} = \int \Delta g_{bulk} dV$, the second term is the domain wall energy, and the third term is the depolarization field energy. In the Landauer model of switching, the domain shape is approximated as half ellipsoid with the small and large axis equal to $r_d$ and $l_d$ correspondingly (Fig. 1a).[16] The domain wall contribution to the free energy in this geometry is approximated as $\Delta G_{wall} = b r_d l_d$, where $b = \sigma_{wall} \pi^2 / 2$ and $\sigma_{wall}$ is the direction-independent domain wall energy. The depolarization energy contribution is $\Delta G_{dep} = c r_d^4 / l_d$, where

$$c = \frac{4\pi P_s^2}{3\varepsilon_{11}} \left[ \ln\left( \frac{2 l_d}{r_d} \sqrt{\frac{\varepsilon_{11}}{\varepsilon_{33}}} \right) - 1 \right] \qquad (3)$$

only weakly depends on the domain geometry.[17]

The mechanism of polarization switching can be analyzed using free energy surfaces representing domain energy as a function of $l_d$, $r_d$. A free energy surface calculated for BaTiO$_3$ ($\sigma$ = 7 mJ/m$^2$, $P_s$ = 0.26 C/m$^2$, $\varepsilon_{11}$ = 2000, $\varepsilon_{33}$ = 120)[18] for the uniform field $E = 10^5$ V/m is shown in Fig. 2a. The free energy surface has a saddle point character and domain grows indefinitely once critical size corresponding to activation barrier for nucleation, $E_a$, is achieved. Minimization of Eq. (2) with respect to $r_d$ and $l_d$ allows the critical domain size and activation energy for nucleation to be estimated as $r_c = \frac{5b}{6a}$, $l_c = \frac{5^{3/2} bc^{1/2}}{6a^{3/2}}$ and $E_a = \frac{5^{5/2} b^3 c^{1/2}}{108 a^{5/2}}$. For the parameters in the text the activation energy for domain nucleation is $E_a = 2.4 \cdot 10^5$ eV for $l_c = 16.4$ µm, $r_c =$ 0.264 μm. Thus, for relatively weak fields corresponding to experimentally measured coercive fields homogeneous domain nucleation is impossible. However, this activation energy is a strong function of the field and for the high fields of order of $10^7$ V/m corresponding to that generated by the tip of radius ~100 nm at potential of 1 V the corresponding parameters are $E_a = 2.17$ eV for $l_c = 11.4$ nm, $r_c = 2.6$ nm. The strong scaling of $E_a$ with bias suggest that even for relatively low tip biases of order of 1-10 V the activation energy becomes small enough to allow thermal fluctuations to overcome the activation barrier resulting in ferroelectric domain formation below the tip. In this case, domain nucleation does not require impurity or other similar nucleation center as is the case for uniform field. Rather, the tip acts as a nucleation center. This behavior can be addresses in a more quantitative way as follows:

In PFM, the electroelastic field distribution below the tip is highly non-uniform and the corresponding domain free energy is

$$\Delta G_{bulk} = \int_V \Delta g_{bulk}(\vec{r}) dV = 2\pi \int_0^{l_d} dz \int_0^{r(z)} \Delta g_{bulk}(r,z) r dr, \qquad (4)$$

where $r(z) = r_d \sqrt{1 - z^2/l_d^2}$. Initial insight into the PFM switching phenomena can be obtained using point charge models applicable for domain sizes $l_d, r_d \gg R, a$, where $R$ is tip radius and $a$ is the contact radius (Fig. 1b), provided that the singularity in the origin is weak enough to ensure the convergence of the integral in Eq. (4). For ferroelectric switching induced by a point charge, $q_s$, located on the surface, $\Delta G_{bulk} = d r_d l_d / (l_d + \gamma r_d)$, where $d = 2 P_s q_s / (\varepsilon_0 + \sqrt{\varepsilon_{11} \varepsilon_{33}})$ and $\gamma = \sqrt{\varepsilon_{33}/\varepsilon_{11}}$. The free energy surface for $q_s = 100$ $e^-$ is illustrated in Fig. 2b. Tip-induced domain switching can be compared to the Landau theory of phase transitions in which domain size is an order parameter. In the case of the point charge on the surface, domain formation is a second order phase transition, since in the vicinity of the point charge the electrostatic field is infinitely large and nucleation always occurs. Similar behavior is expected for a point charge inside ferroelectric material.

For a point charge, $q_a$, located at height, $h$, above the surface, the field in a ferroelectric is finite and nucleation is now a first order phase transition. Examples for different point charge magnitudes and $h = 10$ nm are shown in Figs. 2c,d,e. For $q_a = 100$ e$^-$, the free energy is positive for all $l_d$, $r_d$ and domain doesn't form, for $q_a = 200$ e$^-$ the free energy surface develops a kink. Finally, for $q_a = 400$ e$^-$ the free energy minimum corresponding to the stable domain and the saddle point corresponding to the activation energy for nucleation are clearly seen. This behavior closely resembles the free energy behavior in the first-order phase transition. For large charge magnitudes or small charge-surface separations, the activation energy becomes small, and the free energy surface resembles that for a point charge on the surface.

This analysis can be extended to spherical tip geometry by modeling the tip with a distribution of image charges. For the weak indentation regime (contact radius $a = 0$), the field distribution was derived in Ref. [15] using the image charge method.[19,20] The image charge distribution in the tip can be represented by the set of image charges $Q_i$ located at distances $r_i$ from the center of the sphere such that:

$$Q_{i+1} = \frac{\kappa - 1}{\kappa + 1} \frac{R}{2(R+d) - r_i} Q_i, \quad r_{i+1} = \frac{R^2}{2(R+d) - r_i}, \tag{5a,b}$$

where $R$ is the tip radius, $d$ is the tip-surface separation, $Q_0 = 4\pi\varepsilon_0 RV$, $r_0 = 0$ and $V$ is the tip bias. The tip-surface capacitance is $C_d(d,\kappa)V = \sum_{i=0}^{\infty} Q_i$ and for the conductive tip-dielectric surface

$$C_d = 4\pi\varepsilon_0 R \sinh\beta_0 \sum_{n=1}^{\infty} \left(\frac{\kappa - 1}{\kappa + 1}\right)^{n-1} (\sinh n\beta_0)^{-1}, \tag{6}$$

where $\beta_0 = \text{arccosh}((R+d)/R)$. In the limit of small tip-surface separation, $C_d$ converges to the universal "dielectric" limit.[21] For conductive surfaces, $\kappa \to \infty$, capacitance diverges logarithmically. Potential and field distributions inside the dielectric material can be found using a modified image-charge model as described by Mele:[22]

$$V_i(\rho,z) = \frac{Q_i}{2\pi\varepsilon_0(\kappa+1)} \frac{1}{\sqrt{\rho^2 + (r_i + z/\gamma - d - R)^2}}, \qquad (7)$$

where $\gamma = \sqrt{\kappa_{33}/\kappa_{11}}$ and $\rho$ is radial coordinate along the surface. The total potential inside ferroelectric in the image-charge model is

$$V_{ic}(\rho,z) = \sum_{i=0}^{\infty} V_i. \qquad (8)$$

Far from the contact area, $\rho, z \gg R$, the potential distribution is similar to that generated by a point charge $Q = C_d V$ on the anisotropic dielectric surface:

$$V_{ic}(\rho,z) = \frac{C_d V}{2\pi\varepsilon_0(\kappa+1)} \frac{1}{\sqrt{\rho^2 + (z/\gamma)^2}}. \qquad (9)$$

A similar approximation was used in Ref. [8] to describe the domain switching processes for the domain size larger than the tip radius. For small separations from the contact area, the point-charge approximation is no longer valid and a full description using Eqs. (7,8) is required. A simplified description of the fields inside the material far from the tip-surface junction can still be obtained using an image-charge model of charge $Q = C_d V$ located at distance $h$ from the surface. Simple analysis by Eqs. (5 a,b) indicates that the potential is dominated by the image charges located close the dielectric surface. The cross-over from exact sphere-plane to asymptotic point charge behavior occurs at distances comparable to the tip radius. Given the characteristic size of the tip of order of 10 – 200 nm, a rigorous description of the early stages of polarization switching phenomena in the weak indentation limit necessitates the use of Eq. (8). This is particularly the case for applications such as ultrahigh density ferroelectric recording or ferroelectric nanolithography in thin films, in which minimum achievable domain size (radius ~ 20 nm)[5] is comparable to tip radius of curvature.

The free energy surface calculated using Eq. (7,8) for a tip radius $R = 50$ nm and bias $V = 5$ V has been plotted as shown in Fig. 2f. Similar to the point charge above the surface, domain nucleation is possible only above a threshold tip bias. The bias dependence of domain energy (corresponding to minimum in Figs. 2 b,e,f) and the equilibrium lateral domain size, $r_d$, for BaTiO$_3$ in a point charge model with realistic tip geometry is shown in Fig. 3a,b. The behavior for a spherical tip resembles that for a point charge at ~1 nm separation, as opposed to charge located at 50 nm from the surface at the radius of curvature of the tip, reflecting the concentration of image charges at the tip-surface junction expected for a high-$k$ material. This effective separation and hence critical nucleation bias will be larger for materials with lower dielectric constant such as LiNbO$_3$. At the same time, domain size and energy plotted as a function of tip charge, $Q_{tip} = C_d V_{tip}$ are nearly independent of effective tip radius (not shown), suggesting that the effective tip charge is indeed the parameter that defines the mechanism of switching process.

Arguably, the point charge model provides an oversimplified description of the tip-induced fields. Particularly, the integral Eq. (4) converges only for $\alpha < 2$, where α describes the asymptotic behavior for potential and strain in the functional form $f = x^{-\alpha}$, where $x$ is the distance from tip-surface contact. For high order ferroelectric switching both electrostatic and strain fields decay as $1/x^2$; hence $\alpha = 2$ for ferrobielectric, ferrobielastic and ferroelastoelectric switching and the integral Eq. (4) does not converge, necessitating exact structure of the field to be taken into account.

In addition to this limitation, analysis using electrostatic sphere-plane model does not take into account the effects of non-zero contact area and contact area capacitance, $C_{ca} = 4\kappa\varepsilon_0 a$, where $a$ is contact area, which can be comparable to the sphere plane capacitance for large indentation forces typically use din PFM. This contribution can be addressed in a straightforward manner, however, any electrostatic model based on point charge solution, sphere plane or even more complex solution taking into account contact area effect neglects the electromechanical contribution to the potential inherent in the ferroelectric material.

To extend analysis of ferroelectric polarization switching to a realistic tip geometry including the effect of contact area ($a > 0$) and elastic stress ($P > 0$) and which can also be extended to strain-dependent ferroelectric phenomena, here we use exact field structure for the strong indentation case derived in Ref. [15]. The total potential induced by the tip can be represented as a sum of electrostatic contribution due to the tip bias and electromechanical contribution due to the load force and electromechanical coupling in the material as $\psi = \psi_{el} + \psi_{em}$, where electrostatic contribution is

$$\psi_{el} = -\frac{2\psi_0 H^*}{\pi} \sum_{j=1}^{3} \frac{k_j^*}{\gamma_j^*} \left(N_j^* C_3^* + L_j^* C_4^*\right) \arcsin\left(\frac{a}{l_{2j}}\right) \tag{10}$$

and electromechanical contribution is

$$\psi_{em} = -\frac{H^*}{\pi R} \sum_{j=1}^{3} \frac{k_j^*}{\gamma_j^*} \left(N_j^* C_1^* + L_j^* C_2^*\right) \left[\left(2a^2 + 2z_j^2 - \rho^2\right) \arcsin\left(\frac{l_{1j}}{\rho}\right) + \frac{\left(3l_{1j}^2 - 2a^2\right)\left(l_{2j}^2 - a^2\right)^{1/2}}{a}\right] \tag{11}$$

where corresponding constants are defined in Ref. [15].

Here the calculations are performed numerically for $R$ = 50 nm and $a$ = 3 nm, which corresponds to an indentation force $P$ = 92 nN for BaTiO$_3$. To quantify the domain switching behavior, the free energy density Eq. (1) is calculated using exact formulae for electrostatic field, while the bulk contribution to the free energy Eq. (4) and minimum of the domain free energy as a function of $l_d$, $r_d$ are calculated numerically.[23] The bias dependence of domain energy and lateral domain size is illustrated in Fig. 3 c,d. Note that for large biases, switching behavior is well approximated by a point charge model where the charge magnitude is now related to the indentation parameters by the stiffness relation

$$Q = \frac{4a^3 C_2^*}{3\pi R} + \frac{2a\psi_0 C_4^*}{\pi}, \qquad (12)$$

where $C_3 =$ 15.40 N/Vm and $C_4 =$ 48.54 $10^{-9}$ C/mV for BaTiO$_3$. This is an extension of electrostatic model that takes into account the electromechanical coupling in the material.

For small biases, switching behavior is qualitatively similar to a point charge above a ferroelectric surface and domains can nucleate only above a critical voltage. For BaTiO$_3$, this minimum is calculated to be $V > 0.1$ V, with the minimum domain size being ~ 3 nm, i.e. comparable to the contact radius. Tip flattening during imaging will result in the increase of the contact radius and hence minimal achievable domain size. On the contrary, for small contact area switching is dominated by the field produced by the spherical part of the tip, resulting in optimal conditions for the domain nucleation as reported elsewhere.[24]

To summarize, tip-induced nanoscale ferroelectric switching in weak and strong indentation limits is analyzed using exact electrostatic and electroelastic solutions. It is proposed that domain nucleation can be mapped on the Landau theory of phase transition where domain size is an order parameter and applied bias plays the role of temperature. For a point charge on the surface or inside the ferroelectric, ferroelectric nucleation can be considered as a second order phase transition, while for a charge above the surface and for a realistic tip shape switching is of the first order. In ferroelectric switching for large domain size, the domain size is independent of the contact area and is determined solely by the tip charge or force. At the same time, at the early stages of switching domain size sensitively depends on contact area, which is critically important for high-resolution ferroelectric lithography. Similar analysis can be extended to other modes of polarization switching including ferroelastic and ferroelectroelastic to be reported elsewhere.[24]

**ACKNOWLEDGEMENTS**

Research was performed as a Eugene P. Wigner Fellow and staff member at the Oak Ridge National Laboratory, managed by UT-Battelle, LLC, for the U.S. Department of Energy under Contract DE-AC05-00OR22725 (SVK). AG acknowledges financial support of the National Science Foundation (Grant No. DMR02-35632). The authors thank Prof. E. Ward Plummer (Univ. Tennessee and ORNL) for valuable discussions.

# Figure Captions

**Figure 1.** (a) Domain geometry during tip-induced switching and (b) geometric parameters of tip-surface junction.

**Figure 2.** Free energy surface for domain switching in (a) a uniform field and the field produced by a point charge (b) on the surface and (c,d,e) above the surface. Relevant parameters are given in the text. (f) The free energy surface for a realistic tip shape. Plotted is logarithm of the absolute value of the energy in eV. Solid lines separate regions of opposite signs, indicated by the plus and minus signs. Shown are saddle points (○) and local minima (●).

**Figure 3.** (a) Minimum domain energy and (b) lateral domain size for a point charge on the surface (solid), a point charge at 3 nm (dash), 10 nm (dash dot) and 30 nm (dot) above the surface and for a spherical tip shape (▲) as a function of effective tip charge and tip bias. The positions 1, 2, and 3 for $h = 10$ nm correspond to free energy surfaces in Fig. 2 c,d,e correspondingly. Minimum domain energy (c) and lateral domain size (d) as a function of the effective tip charge and tip bias for parameters given in the text. Solid line corresponds to point charge model, (■)- numerical solution for ferroelectric switching in a strong indentation regime.

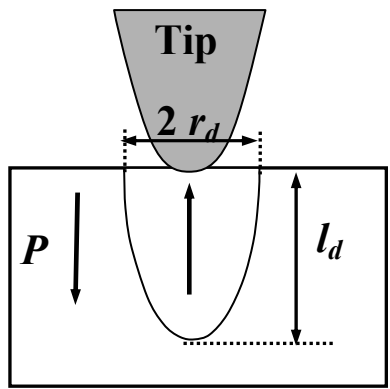 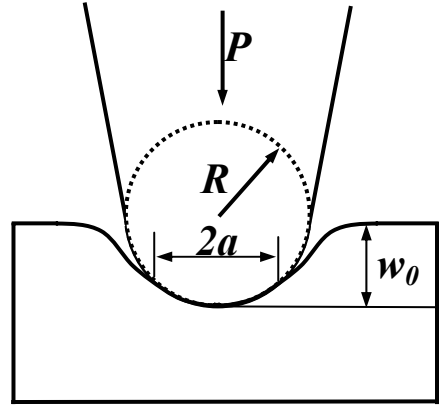

**Fig. 1.** S.V. Kalinin, A. Gruverman, J. Shin *et al*.

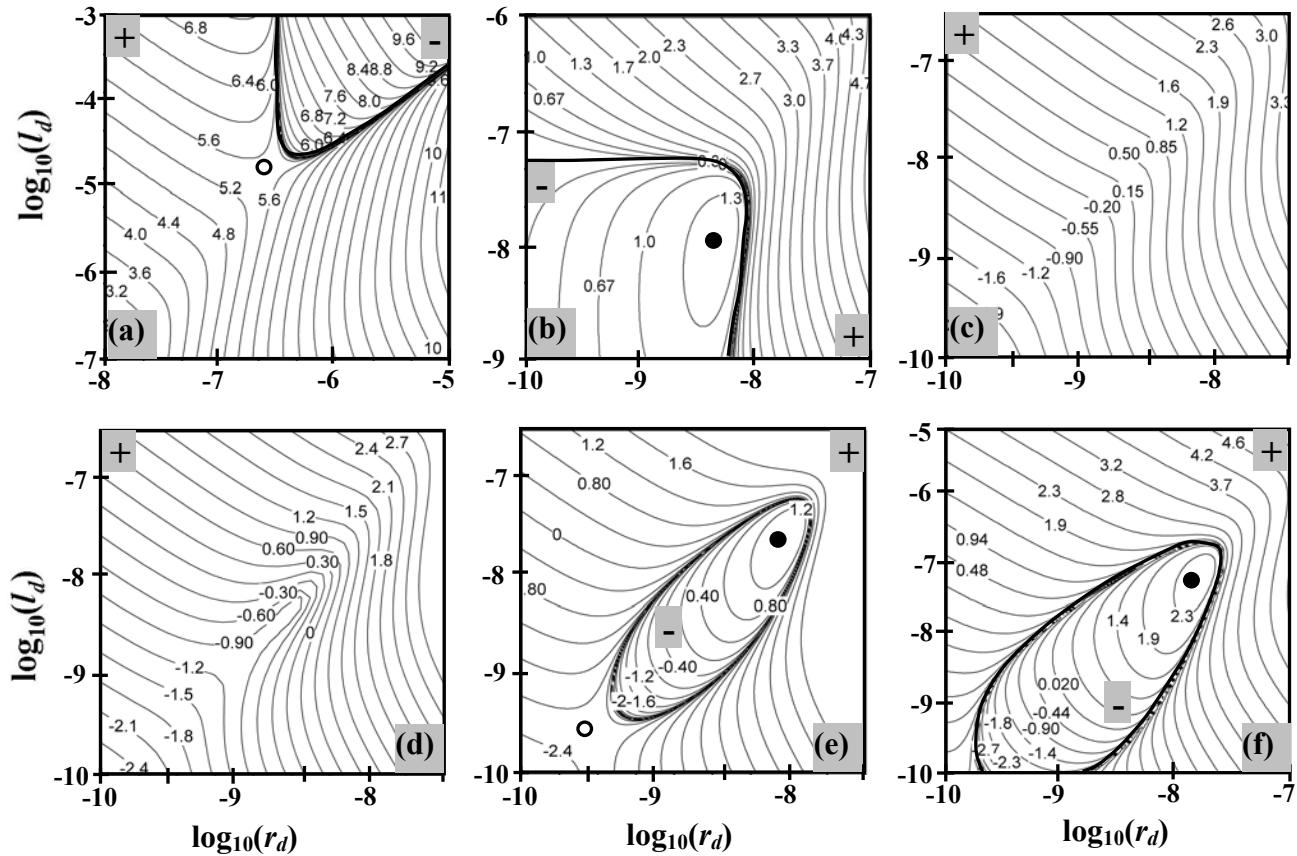

**Fig. 2.** S.V. Kalinin, A. Gruverman, J. Shin *et al*.

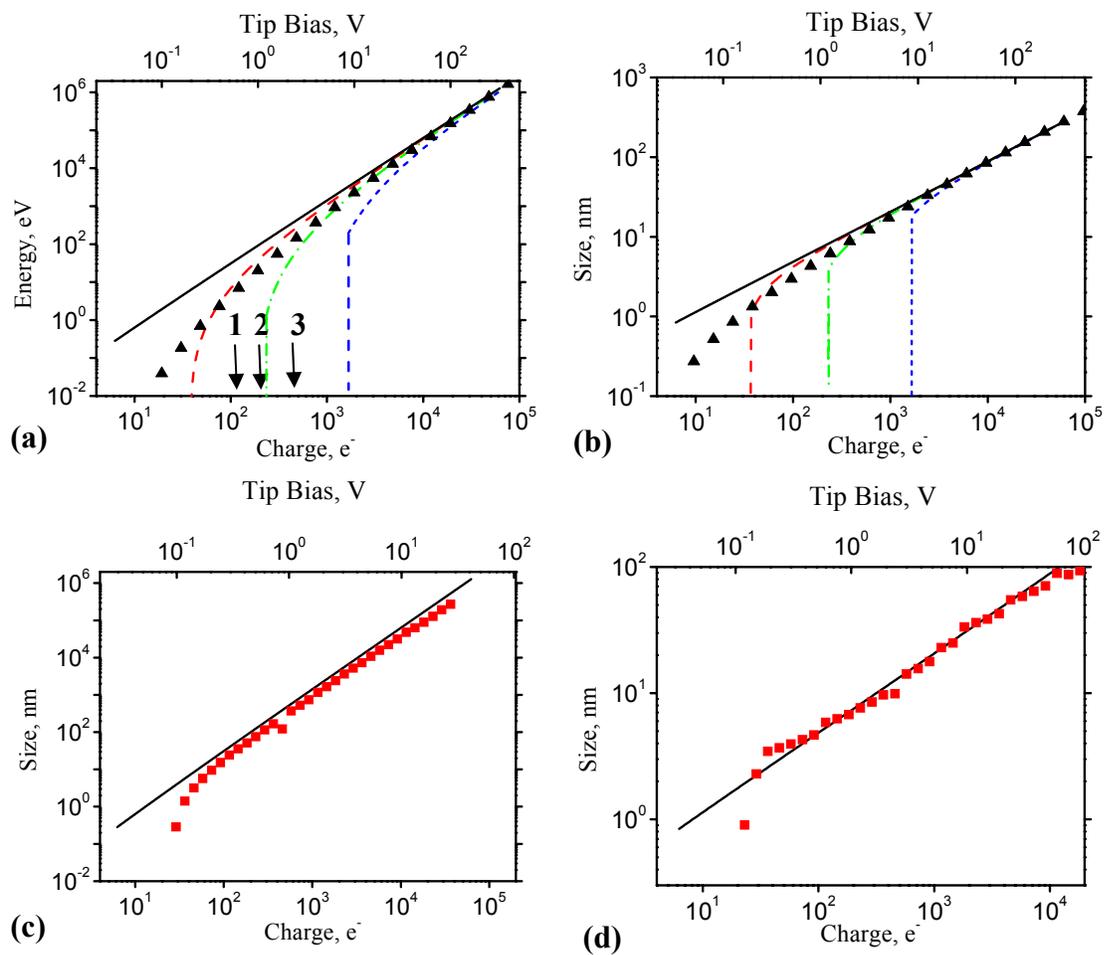

**Fig. 3.** S.V. Kalinin, A. Gruverman, J. Shin *et al*.